\documentclass[10pt,letterpaper]{article}
\usepackage[letterpaper,margin=20mm]{geometry}

\usepackage{amsmath}
\usepackage{xcolor}
\usepackage{wasysym} 
\usepackage{subcaption} 
\usepackage{comment}
\usepackage{graphicx} 
\usepackage{cite}
\usepackage{url}

\newcommand{\AG}[1]{\textcolor{black}{#1}}    

\usepackage{hyperref}
\raggedright
\begin{document}
\vspace*{0.2in}

{\Large

\textbf\newline{{Dynamical Model for the Sustainable Development Goals }}}\\


Alberto Garc\'ia-Rodr\'iguez\textsuperscript{1,4,5,6}, 
Tzipe Govezensky\textsuperscript{2,*},
Julia Tagüeña\textsuperscript{3,6},
Kimmo K.Kaski\textsuperscript{4},
Rafael A. Barrio\textsuperscript{1}
\bigskip

\textbf{1} Instituto de Física, Universidad Nacional Autónoma de México, Coyoac\'an, CDMX 04510, México
\\
\textbf{2} Instituto de Investigaciones Biomédicas, Universidad Nacional Autónoma de México, Coyoac\'an, CDMX 04510, México
\\
\textbf{3} Instituto de Energ\'ias Renovables (IER), Universidad Nacional Aut\'onoma de M\'exico, Temixco, Morelos, M\'exico (Mexico)
\\
\textbf{4} Department of Computer Science, Aalto University School of Science, Helsinki, Finland
\\
\textbf{5} Instituto de Investigaciones en Matem\'aticas Aplicadas y en Sistemas, Universidad Nacional Aut\'onoma de M\'exico (UNAM), CDMX, Mexico.
\\
\textbf{6} Centro de Ciencias de la Complejidad, Universidad Nacional Autónoma de México, Coyoac\'an, CDMX, México

\bigskip


* tzipeg@hotmail.com

\section*{Abstract}
    
The 2030 Agenda for Sustainable Development of the United Nations outlines 17 goals~\cite{URL:Goals} \AG{as global challenges} for countries of the world to address 
in their development. 
However, the progress of countries towards these goals has been \AG{much} slower than expected.
In a previous study~\cite{SDG2025}, we analyzed the data over two decades (2000–2022), using unsupervised machine learning techniques. Based on this study, we take into account three main factors to construct a mathematical model to simulate and predict the dynamical behavior of the SDGs. These factors are: 1) the distribution of amount of resources that each country uses to meet the goals, 2) the cooperation between countries, and 3) the correlations between the goals. In this work, we show that the model is capable of reproducing the real data and therefore could be used to simulate hypothetical scenarios that could help to improve 
actions towards optimal fulfillment of the goals.


\section{Introduction}
{\color{black} As we approach 2030, the target year for reaching the goals of United Nations Sustainable Development agenda set in 2015, 
it has become quite clear through recent analysis studies \cite{SDG2025,su17167411} that these ambitious goals 
will not be achieved}. 
 {\color{black}In this situation and to better understand the dynamics of sustainable development goals, a mathematical model is needed. This is important especially to gain insight into the performance of different Sustainable Development Goals (SDGs) due to}  
complex interactions between social, economic, environmental, and institutional factors, which can be considered as the four columns of sustainability. {\color{black}Furthermore, modelling can serve \AG{the role of} 
"what-if" experiments for forecasting different future scenarios and testing strategies for achieving the SDGs, as well as being able to optimize resources, for instance, to mitigate climate change. Although not in the scope of the present study, these models could provide policy-makers tools to include synergies and trade-offs between 
the SDGs and evidence to design policies.} 
%

{\color{black}In the present study our goal} 
is not only to combine empirical data and 
develop a toolbox such as the Agenda 2030 Compass \cite{compass}, but also to reach a deeper understanding of the causes underneath. This Compass identifies synergies and trade-offs between the SDGs and was developed \AG{in} 
collaboration between 
industry, public authorities,
researchers, and civil society; originally \AG{it was} developed as a spin-off of 
the Swedish steel industry’s 2025 vision (Stockholm Environment Institute). \AG{For describing the behavior of SDGs there} 
are a couple of dynamical models: the iSDG Model~\cite{isdg} and the Functional Enviro-economic Linkages Integrated Nexus (FeliX) Model~\cite{redivo_2022_system}. \AG{In addition, the} 
Millennium Institute has developed a 
system dynamics model, the integrated SDG (iSDG), which 
simulates the trends for the SDGs until 2030 for each of the SDG indicators under a business-as-usual scenario. \AG{It} 
supports the analysis of alternative policy scenarios~\cite{isdg}. The iSDG Model has 30 
{\color{black}interdependent} sectors (i.e., economic, social, environmental \AG{which}) 
simulate SDGs until 2030 using specific data \AG{that enable} 
trade-off between goals. The FeliX Model itegrates 10 global systems (i.e. population, energy, climate, etc.) and models \AG{eight (8)} SDGs, emphasizing human-natural system feedbacks. Both models have data gaps and financial and power imbalances.

In a recent PNAS Perspectives~\cite{Selin2023}, \AG{the authors evaluated} 
progress \AG{of sustainable development by} 
modeling nature–society systems. 
They also highlight examples for each of the four stages of modeling practice—defining purpose, selecting components, analyzing interactions, and assessing interventions.  However, their potential has many possibilities for sustainable science that have not yet been fully appreciated. 
There are a few models for specific regions, such as \AG{the one by} Topf et al.~\cite{TOPF2023107645}, \AG{proposing} 
a new index (the Sustainable Development Pathway Index, SDPI) that uses the Euclidean norm to measure the distance between the ideal and the actual development pathways of any geopolitical unit in a determined time frame. They \AG{have} evaluated the development pathways of 517 municipalities in the Brazilian Amazon, from 1991 to 2010, \AG{while at the same time} 
analyzing indicators of ecological and socioeconomic infrastructures. 
\AG{Recently yet another study}~\cite{su152215929} 
developed 
a forecasting model that considered the interactions of the SDGs and simulated their progress 
from 2021 to 2030 for 41 cities in the Yangtze River Delta under various sustainable development paths. The results indicated that the cities with the highest levels of sustainable development in the Yangtze River Delta, \AG{finding that only half of them would achieve} 
the goals by 2030 if they \AG{continue along the present paths}. 
\AG{In additions, they} 
optimized the development pathways taking into account the development costs and goal attainment. 
\AG{In \cite{ALLEN2016199} the authors reviewed a broad range of 80 different quantitative models on the basis of their strengths, weaknesses, and general utility for the purpose of exploring policy scenarios to achieve SDGs at the macroeconomic level}. 
%
 These models are local and simplify reality, and 
 their accuracy is limited \AG{due to} 
 data limitations and inherent uncertainties. \AG{In contrast to this our aim here} 
 is to produce a global \AG{model of SDGs dynamics that is} 
 based on controllable parameters.

 \AG{In several studies it has become evident that in various countries or groups of them the relationships between different SDGs are very complex following different development trajectories due to historical, social, and economic differences. Furthermore, as has been analyzed~\cite{SDG2025}, achieving a certain goal can jeopardize the progress on another goal, which makes  building a descriptive model quite challenging. Then the predictive capabilities of such adynamic model are fundamentally based on its set of parameters, which capture the intricate relationships within socio-economic and environmental systems. These parameters, both physical and socioeconomic, drive simulations and allow for the exploration of future scenarios.}

\section{Model}

In order to model the dynamical behavior of the processes by which countries attack the problem to attain the SDG goal, we \AG{assume} 
that the main ingredient is the amount of resources that they dedicate to follow the  actions needed to solve their particular problems. Our dynamical variable should be a number that reflects an index that measures the progress to attain each goal as a function of time. Let us name it $x(i,\alpha, t)$, with the index $i=[1,N]$ indicating a country, and the index $\alpha=[1,N_g]$ label a goal, where the maximum value for $N_g$ is $17$ based on the 17 SDG's. One has to take into account the interactions between different goals. In addition, we recognize the fact that past events influence the present actions taken. 

\AG{To describe the complex progress of an SDG of a country, $x(i,\alpha, t)$, we propose} 
the following dynamical equation 

\begin{equation}
\begin{aligned}
    \frac{x(i,\alpha,t)}{dt} &= \{r(i) * r_{0} * (1 - e(i,\alpha)\}*[1-x(i,\alpha)]/g_l +  g_k*nn(i,\alpha,t) + \\ 
    & \frac{g_k}{2}* nnn(i,\alpha,t) + +g_{int}(i,\alpha,t) + \frac{1}{m}\int_{t-m}^{t}x(i,\alpha,t')dt'.
    \end{aligned}
    \label{dyn}
\end{equation}


Let us explain in detail the meaning of each term in Eq \ref{dyn}. The first term on the right represents the progress made in terms of the resources  $r(i)$ that  each country allocates to carry out the actions  and the  percentage of available resources $r_0$ spent over a given period of time. The normalized matrix $e(i,\alpha)$ encodes the amount that each country is spending on each SDG goal. The adjustable parameter $g_l$ sets the limit value of the variable $x$ to achieve the goal. 
In the second term on the right we take into account the interactions between the nearest neighbor countries in a network defined by an $N \times N$ adjacency matrix $A$ as follows 
\[nn(i,\alpha,t)=\frac{1}{k_i}\sum_{j=1}^{k_i} A(i,j)x(j,\alpha,t),\]
where $j$ is a neighbor of $i$ and $k_i$ is the degree. 
\AG{Then the third term on the right of Eq \ref{dyn}}, $nnn(i,j,t)$, is the sum of the scores of the next nearest neighbors, as found in $A(i,j)^2$ of $i$ in the $\alpha$ goal. Here, $g_k$ is an adjustable parameter that weights the influence of neighbors on the network. 
The fourth term \AG{is calculated from Eq \ref{dyn2}} 

\begin{equation}
\begin{aligned}
        \frac{g_{int}(i,\alpha,t)}{dt}& = g_i(i)*(gl_{inks}(i,\alpha) * x(i,\alpha,t)) .
    \end{aligned}
    \label{dyn2}
\end{equation}
\AG{in which} the $[N_g,N_g]$ entries of $g_{links}$ describe 
the  correlations between \AG{the ($N_g = 17$)} SDG goals,  taking into account the fact that progress in some action could improve progress in others, if the correlation is positive or jeopardize others with a negative correlation. This matrix dictates the dynamics of the term $g_{int}$, which is coupled to the dynamics of \AG{progress indicator}, $x(i,\alpha,t)$. This term is weighted by the vector $g_i(i)$, which regulates the impact of the correlations between the goals of each country. The last term of Eq. \ref{dyn} is the memory that influences the progress of the actions.\\

\section{Numerical integration}
\AG{To solve the time dependent} Equations \ref{dyn} and \ref{dyn2} \AG{we integrate them} 
numerically using the 
Euler method with a time step of $dt=0.001$, which is small enough to avoid artifacts and instabilities, and 
can be adjusted to represent real time when fitting to real data.
%
%

Next, we present some example calculations \AG{to demonstrate} the versatility of the results obtained by varying the important factors, namely, $A$ and $g_i$ and also by using  random numbers, $x_0(i)$, for the initial values of the progress made by each country. \AG{Here, we find that variations} 
in the magnitude of resource investment significantly influence the rate at which countries progress toward achieving their goals,  coefficient $g_k$ regulates the relative importance of neighbor influence and the importance of correlation interactions is regulated by $g_i$.

\begin{figure}[ht!]
\begin{center}    
    \includegraphics[width=.6\linewidth]{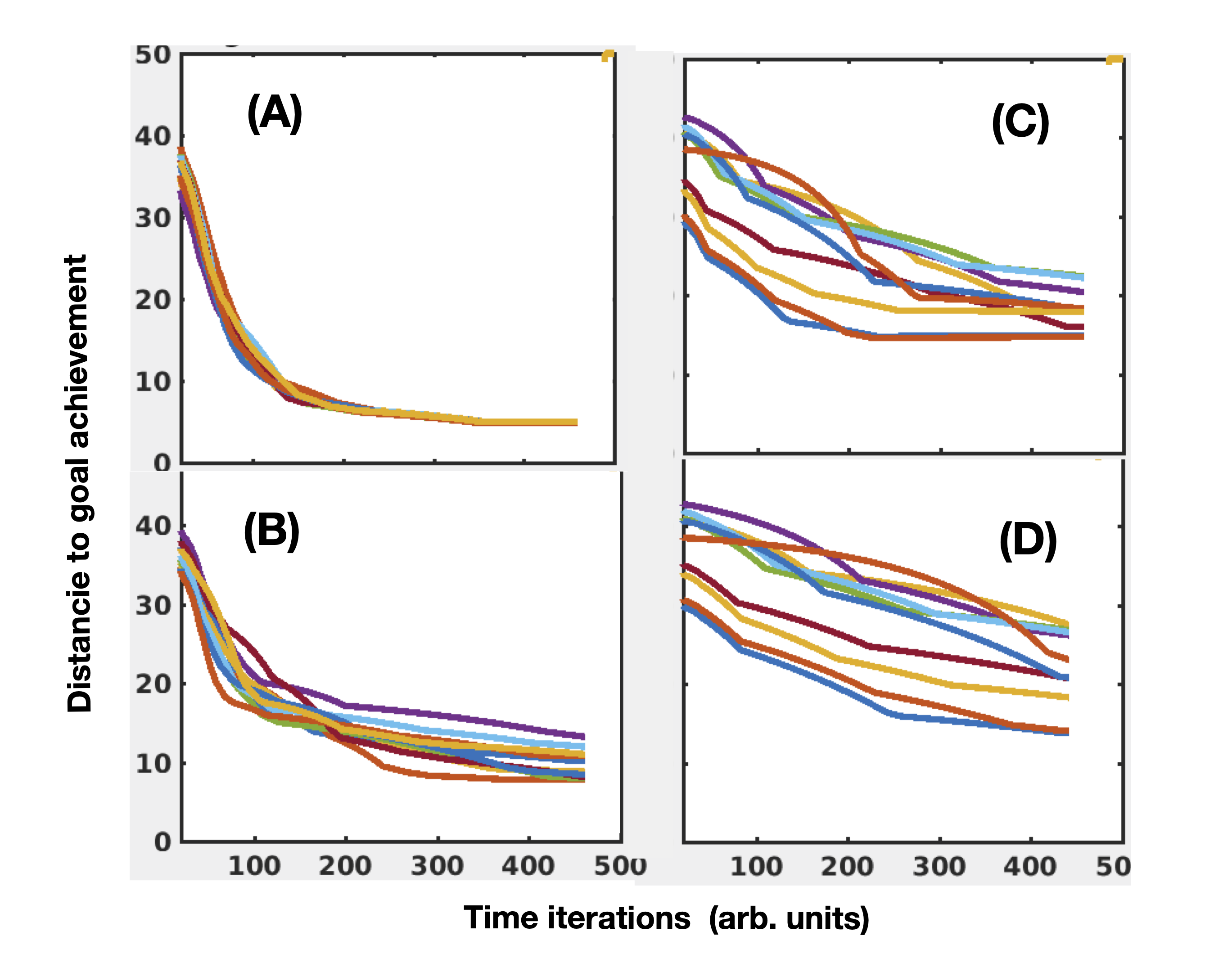}
    \caption{(A) calculation when the neighbor matrix is complete, (B) when neighbor matrix is empty. (C) Calculation with $g_i=8$. (D) $g_i=4$ for all countries. }
    \label{fig:fig1}
    \end{center}
\end{figure}

In Fig.\ref{fig:fig1} we show some examples of the results obtained varying the adjacency matrix ($A$) and the impact of correlations between goals ($g_i$). The parameters used in all calculations were: Number of countries=10, Number of goals=6, $r$, which represents the resources available to each country, is a vector whose entries were 10 random numbers, $u$, between 30 and 100 following a power law $u^{-1/2}$, that is $r=[30.43,33.07,39.21,31.29,32.32,30.00,30.02,30.35,39.67,100.00]$, $g_l=50$,  $g_k=0.5$ and $r_0=0.1$.

\
In Fig.\ref{fig:fig1} (A) we notice that when all countries cooperate, the behavior of all individual countries is quite similar and that they obtain their goals in the shortest time. The fact that no country reaches the zero mark is because the correlation matrix used is a random matrix with many goals being anti correlated. This can be compared with  Fig.\ref{fig:fig1} (B), a result obtained when the neighbor matrices are diagonal (i.e., no cooperation).  Observe the large dispersion of the curve and the slow progress of all countries. Figs.\ref{fig:fig1}(C) and (D) show irregular behavior, slow progress, and a higher level of stagnation when the importance of the correlations is increased. 

These results indicate that cooperation among countries reduces the differences and could contribute to achieving common goals. The SDG's are undoubtedly correlated and the correlations could not only slow \AG{achieving} the \AG{goals} 
but also produce negative outcomes.


\newpage

\section{
\AG{Testing the model with real data}}

\AG{In this section we 
test the model by applying it to the real SDG data.}
\AG{In order to do that we need to \AG{feed into} 
the model 
quantities extracted from the data. \AG{This means that} 
we have to give values to the adjacency matrix ($A$) entries that reflect the real cooperation between countries, the amount of resources ($r$) spent by each one \AG{of them}, the correlation between goals ( $g_{links})$}.


\begin{figure}[ht!]
 \begin{centering}   
 \includegraphics[width=.8\linewidth]{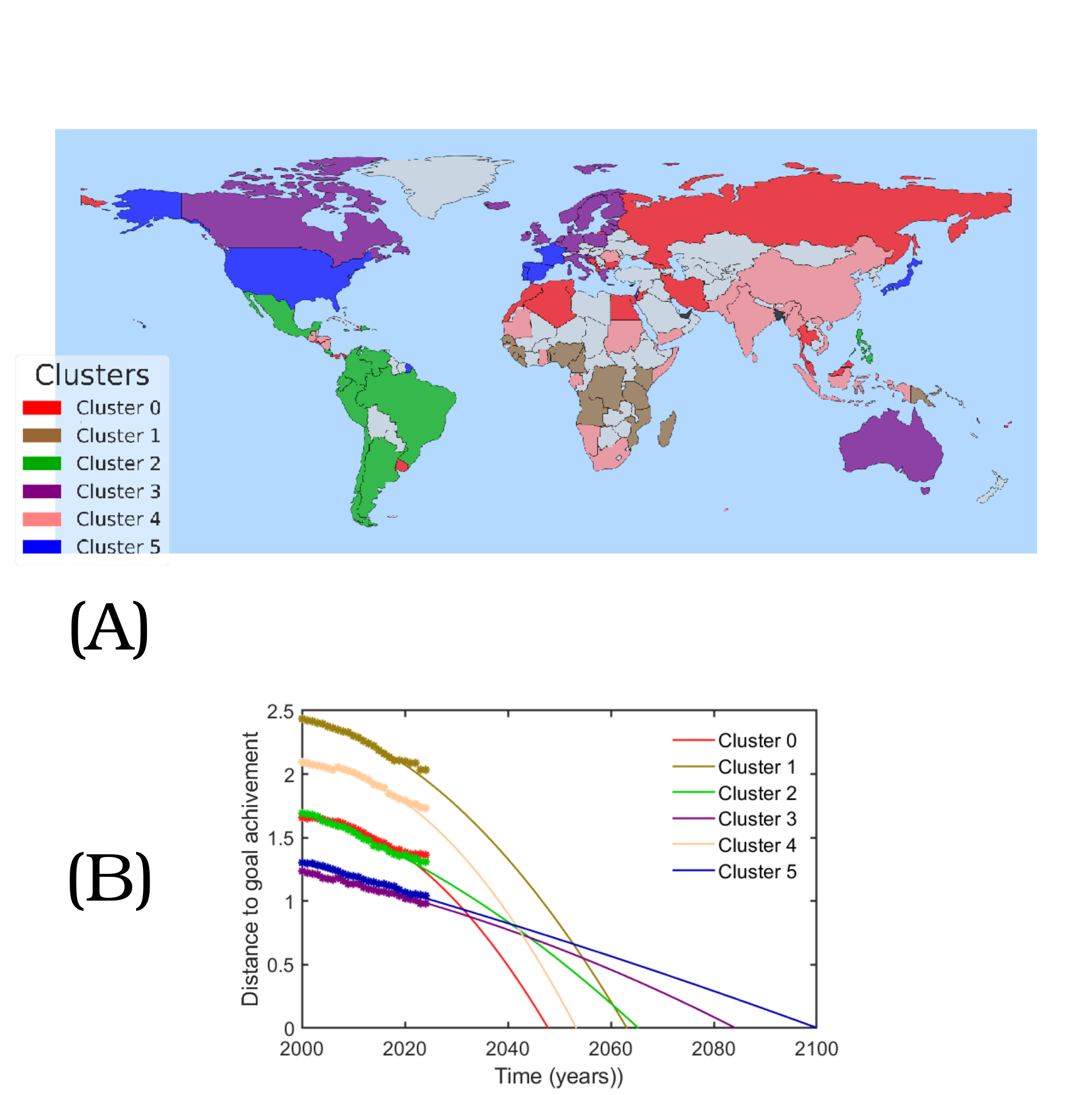}
    \caption{(A) Map showing the grouping of countries in six clusters. (B) Real data shown as "*" and the lines are quadratic extrapolations to predict \AG{each cluster's temporal progress achieving their goals totally}. }
    \label{fig:fig3}
    \end{centering}
\end{figure}

In our previous work (ref. \cite{SDG2025}) we \AG{categorized countries into six similarly behaving groups using unsupervised machine learning \AG{and} clustering technique\AG{s}}. 
Fig.~\ref{fig:fig3}(A) shows \AG{the world} map \AG{with countries colored} 
according to this clustering, and Fig.~\ref{fig:fig3}(B) shows \AG{temporal development of Euclidian distance of these clusters towards completion of every goal, first for years 2000 to 2024 with real SDG data and from thereon with extrapolation using a quadratic function}. 
Although this extrapolation is useful to 
see \AG{how the temporal progress towards the 2030 agenda goals could proceed}, 
the dynamical model proposed in this work is \AG{aimed} 
to get 
\AG{a process-based understanding how the agenda goals could in real situation be achieved}.


\begin{figure}[ht!]
    \includegraphics[width=\textwidth]{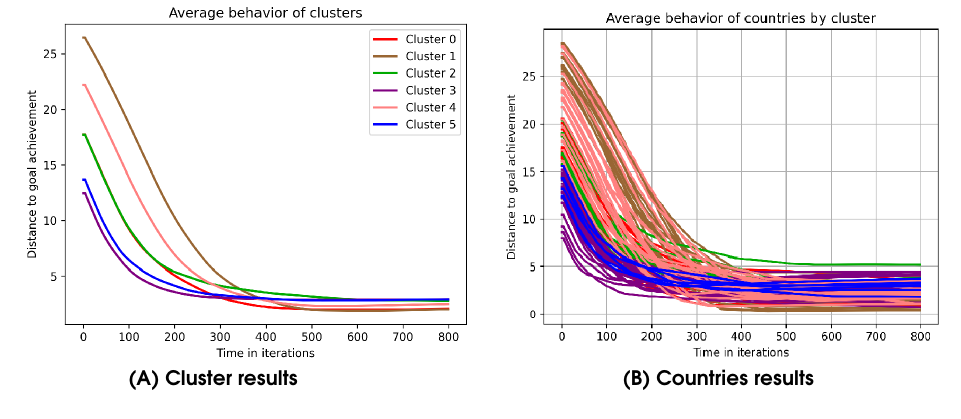}   
        \caption{(A) Results using real data for the clusters shown in Fig.  \ref{fig:fig3} (A) and (B) results simulated for all 88 countries following there cluster color.}
        \label{fig:real_simulation}
\end{figure}

\begin{figure}[ht!]    
\includegraphics[width=\textwidth]{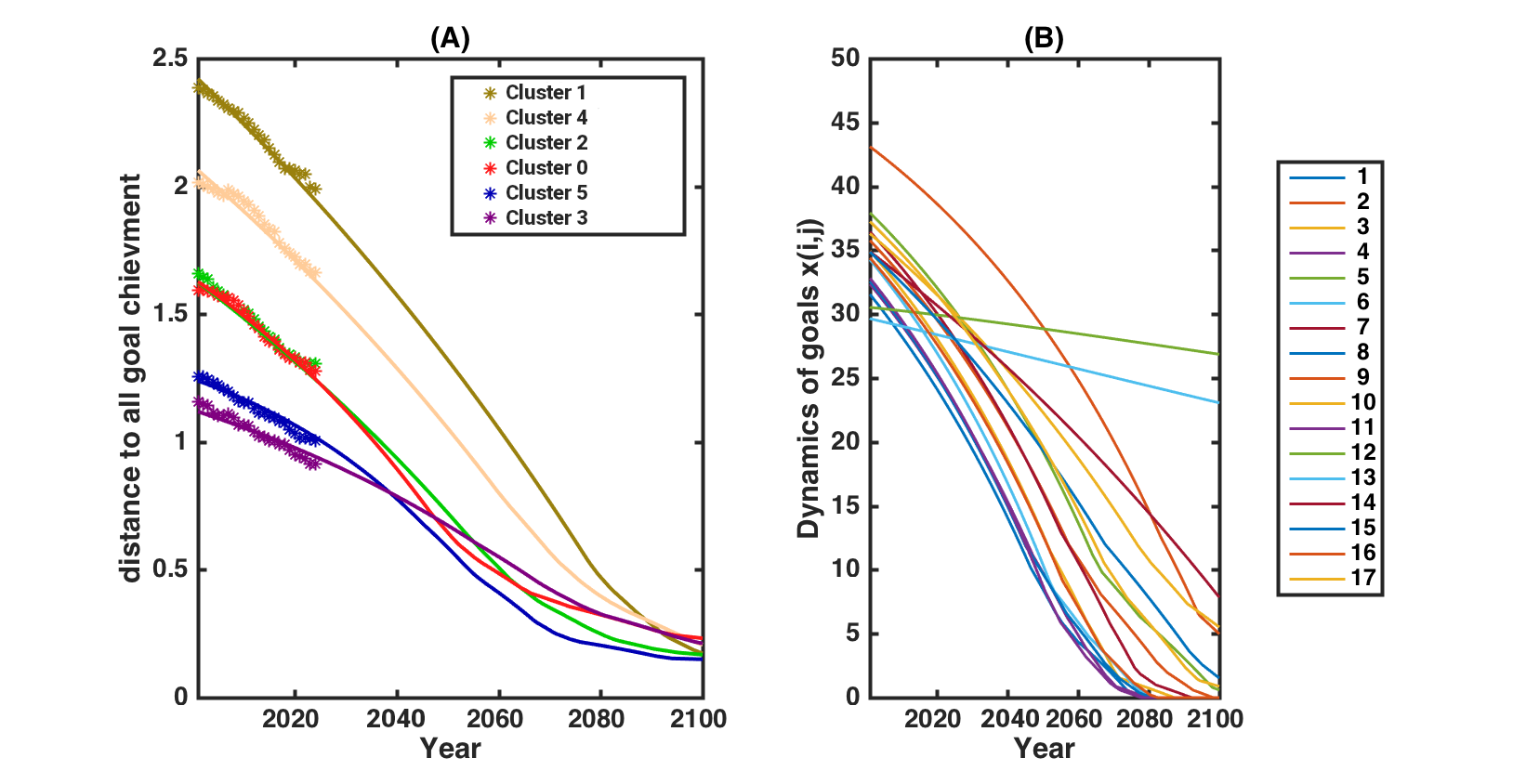}
         \caption{(A) Comparison between the real data (color stars) and the results of the calculation (color lines) using the information provided by the data to adjust the parameters. (B) Dynamical behaviour of the 17 SDG goals. Observe the remarkable behavior of goals 12 and 13, (see Appendix A) which are almost flat due to anti-correlations.}
         \label{fig:fig4}     
\end{figure}

\AG{The results} 
shown in Fig.~\ref{fig:real_simulation} and Fig. \ref{fig:fig4} used the correlation matrix between the goals extracted from the data and for the initial conditions of $x(i, j)$ we used the 2000 SDG scores (ref.~\cite{SDG2025}). In Fig.~\ref{fig:real_simulation}(A) we show the results obtained assuming that we have only six clusters, which have similar SDG patterns (ref. \cite{sdgindex_2024_sustainable}), and 
\AG{thus each cluster could} 
be treated as a single country \AG{and then setting the initial values for its $x(i, j)$ the averages of the 2000 data over the countries belonging to the same cluster}. 
Fig.~\ref{fig:real_simulation}(B) shows the results obtained when using data for single countries, colored according to the cluster to which they belong. The Dynamical behavior of counties of the same cluster are grouped together, showing a very similar behavior observed when modeling the dynamics of clusters. The parameters used in these figures were the number of countries=88, number of goals=17, $g_k=2$, $r_0=0.1$, $g_l=50$, and the full vector $r$ was taken from the published GDP values (see Appendix B).

\AG{Fig.~\ref{fig:real_simulation}(A) shows that countries belonging to the same cluster (same colour) behave in a similar way. Therefore, in order to simplify the calculations, from now on we 
consider the average behaviour over countries in the same cluster.} 
%
Next, we have to adjust \AG{the quantities and parameters of the model 
to faithfully reproduce the data from 2000 to 2024. First we notice from the recorded data in Fig.~\ref{fig:real_simulation}(A) that the 6 clusters, now  considered as countries, group as three pairs. Thus, we used a $6\times6$ adjacency matrix with non zero entries only between pairs. }
%
\AG{Regarding the available resources of these six cluster countries ($r$) we used normalized relative GDP averages over the countries in the cluster as follows 
$r=[80, 160,  240,  320,  400,  480]$.} The annual expenditure was assumed to be 10\% \ of the resources. \AG{As for the correlations between the (Ng = 17)
SDG goals we used the same $g_{links}$ matrix as in our previous work \cite{SDG2025}}.

The results for the best fit to the real data are shown in Fig.\ref{fig:fig4} \AG{for which the} 
relevant parameters used in Eq.\ref{dyn} were $g_i=[0.2,    0.5,    0.3,    0.15,   0.15,    0.15$],  $g_k=0.5$, $g_k=1.5$, and $m=5$. Here a 10 year time span corresponds to 100 iterations with the time step of $dt=10^{-3}$. 
Note that in Fig. \ref{fig:fig4} (B), as we obtained in Ref. \cite{SDG2025}, \AG{the SDG} goals 12 (Responsible consumption and Production) and 13 (Climate Action) \AG{have a slow dynamics since they} are highly positively correlated with each other, yet strongly anti-correlated with nearly all the other goals.

\section{Discussion of possible scenarios}

\subsection{Variations in the magnitude of $g_k$.}

\AG{As our model seems to be able to reproduce the behaviour of real data and to predict the continuation to the future, it would be interesting to see changes in the parameters and predict the behavior of SDG's in various hypothetical conditions, i.e. explore different  scenarious.} 
This will allow us to decide which conditions should be changed to improve the attainment of the goals or which actions should \AG{be taken} 
to attain the goals in a reasonable and feasible time.

\begin{figure}[htb!]         \includegraphics[width=.8\textwidth]{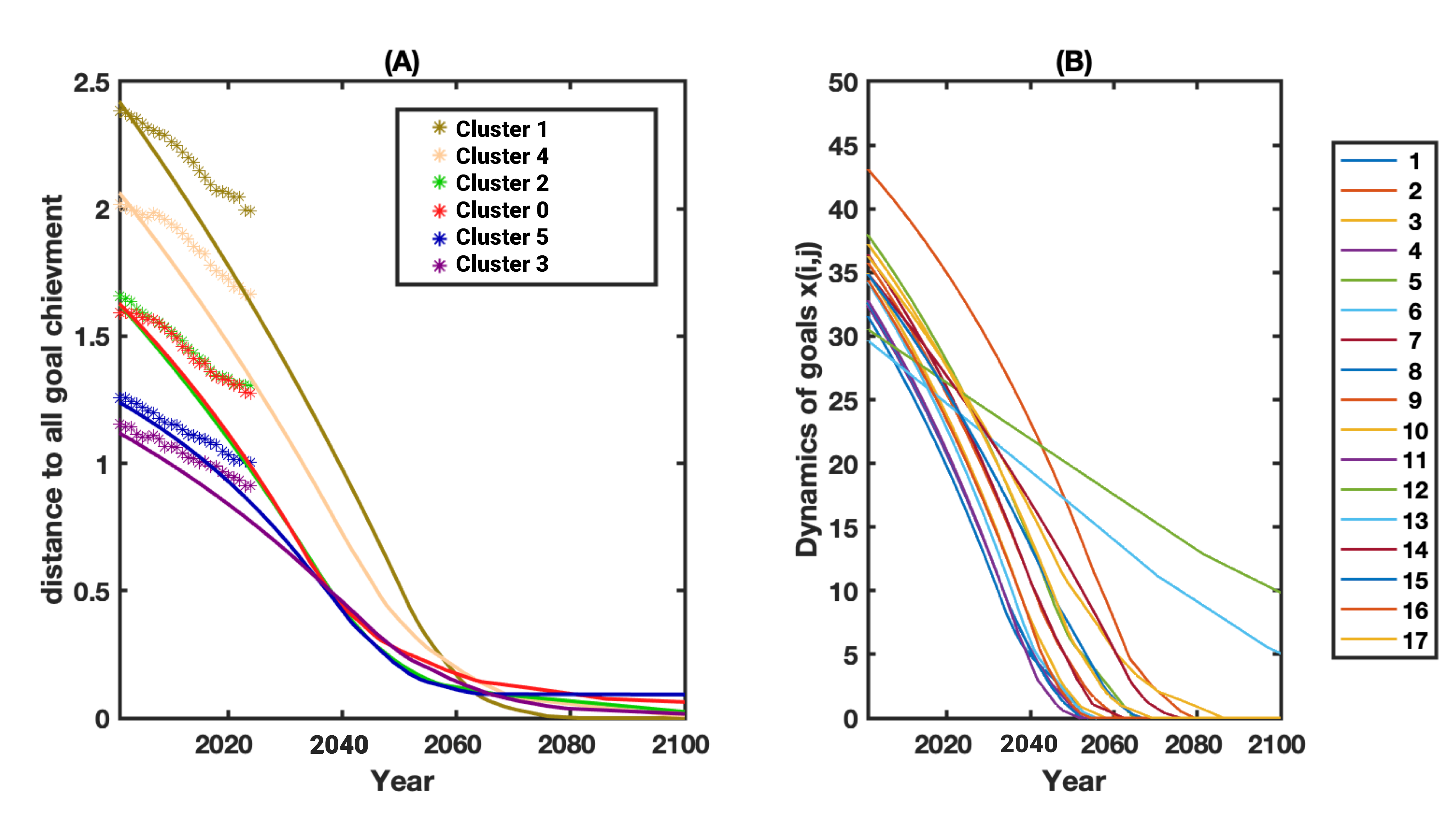}
         \caption{(A) Calculation made with $g_k=10.5$, real data (color stars) and the results of the calculation (color lines). (B) Dynamical behaviour of the 17 goals. Dynamical behaviour of the 17 SDG goals. Observe the remarkable behavior of goals 12 and 13, which now are progressing in spite of the anti correlations.}
         \label{fig:full}
\end{figure}

\AG{Let us} 
start by looking at the role of correlations between \AG{the SDG} goals. \AG{For that we} 
use the correlation matrix obtained from the real data in Ref.~\cite{SDG2025}, and vary the parameter $g_i$  to investigate \AG{whether} 
it is possible to reach all the goals, if ever. If we change the importance of interactions among countries by increasing the value of this parameter from $g_i=$ 1.3 to 10.5, we obtain the results shown in Fig.\ref{fig:full}.
\AG{In this case we observe} 
that the dynamics \AG{for reaching the goals} 
is faster,  but  after year $\approx 2080$ the system becomes stagnant and complete success (i.e., reaching zero) is not achieved. However, by comparing Fig.\ref{fig:full}(B) with Fig.\ref{fig:fig4}(B) we notice that even the goals that are anti-correlated show a monotonic approach to success.
\AG{Hence, we find} 
that the correlation parameter, which reflects cooperation among countries, accelerates the success process, but it reaches a limit due to the correlation matrix.

\subsection{Effecs of correlations between goals}

\begin{figure}[ht!]        \includegraphics[width=.8\textwidth]{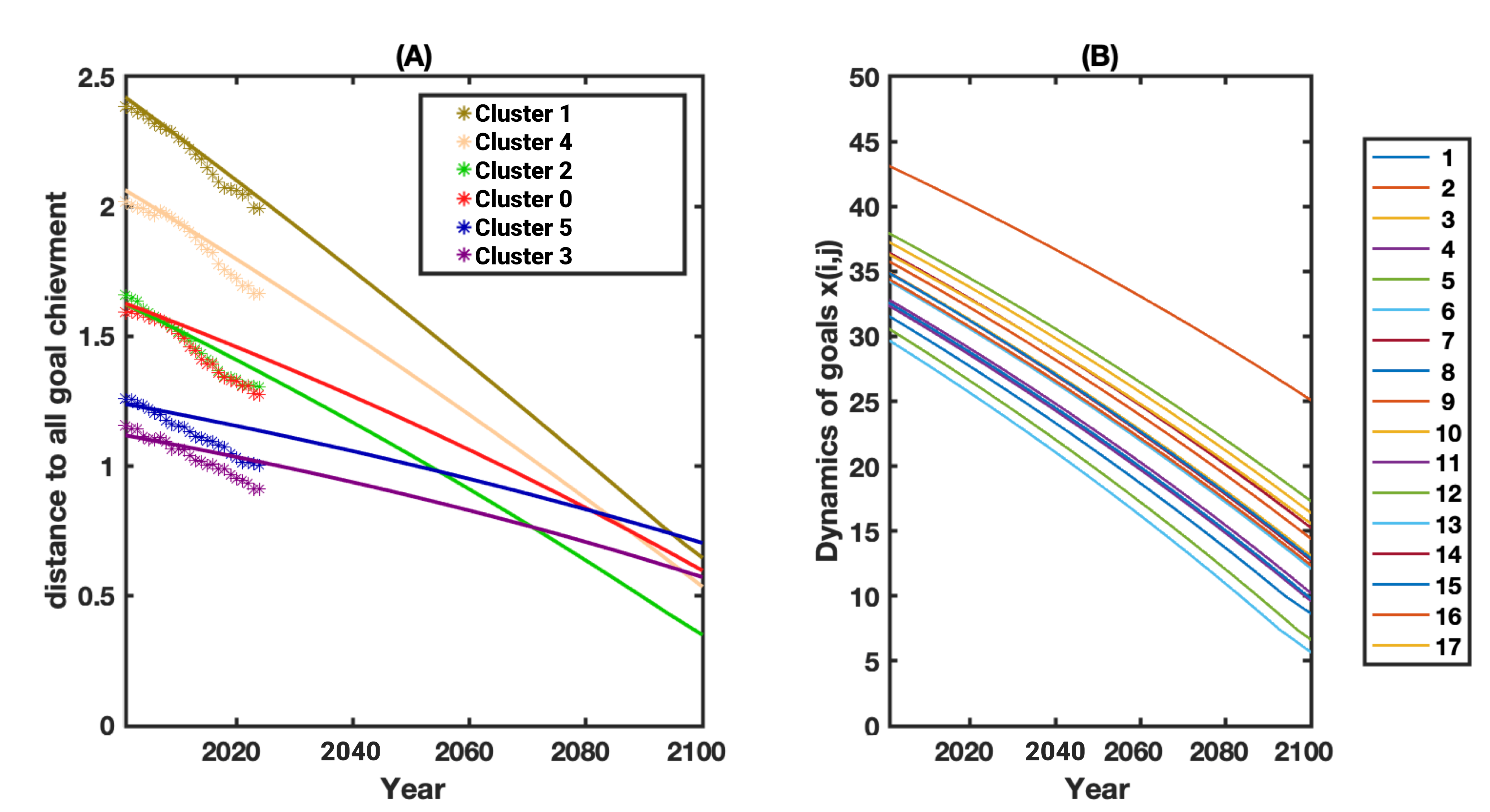}
     \caption{(A)trajectories of clusters when $g_i=0$, meaning that there is no correlation between goals, real data (color stars) and the results of the calculation (color lines). (B) Dynamical behaviour of the 17 SDG goals when they are totally uncorrelated.}
        \label{nocorrelation}
\end{figure}

\begin{figure}[ht!]
         \includegraphics[width=.8\columnwidth]{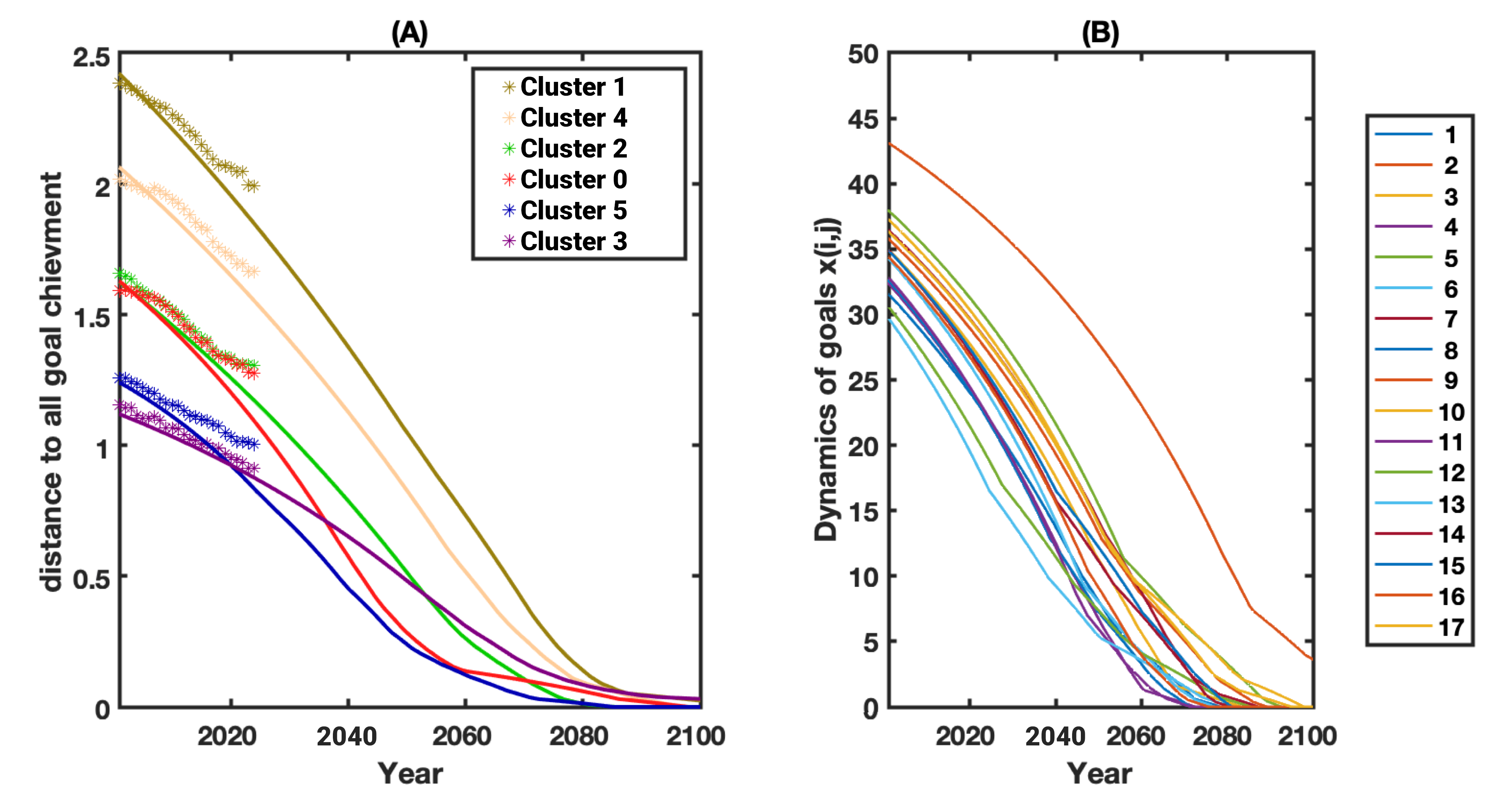}
         \caption{(A) Trajectories of clusters  when $g_i=1$, and all correlations are positive, real data (color stars) and the results of the calculation (color lines). (B) Dynamical behaviour of the 17 SDG goals when all correlations are positive.}
        \label{psocorr}
\end{figure}

From the results exhibited \AG{above}we can conclude that correlations \AG{between SDGs} are important and that it is not possible to reach complete success if one defines the goals as they are presented up to now. 
\AG{Then let} 
us investigate what happens if one chooses goals that are totally uncorrelated. This corresponds to the situation in which $g_i=0$.
In Fig.\ref{nocorrelation} we exhibit the results. Surprisingly, in Fig. \ref{nocorrelation} (A) we see that the dynamics is very much slower and one is not able to attain the goals even in one hundred years. The fact that the goals are not correlated is evident in Fig.\ref{nocorrelation}(B), where we can see that all goals behave roughly in the same way. 
\AG{Then one} 
could conclude that correlations between goals should be present because actions directed to attain a certain goal help achieve other goals \AG{that are} positively correlated.
In order to verify this conclusion we made a calculation in which we used random correlation matrix with all its elements positive and a relatively small strength ($g_i=1$), this is shown in Fig.\ref{psocorr}. \AG{Thefore, it is clear that positive correlation between SDGs would eventually lead to success}.

\subsection{Effect of economic resources}

\begin{figure}[ht!]         \includegraphics[width=1\textwidth]{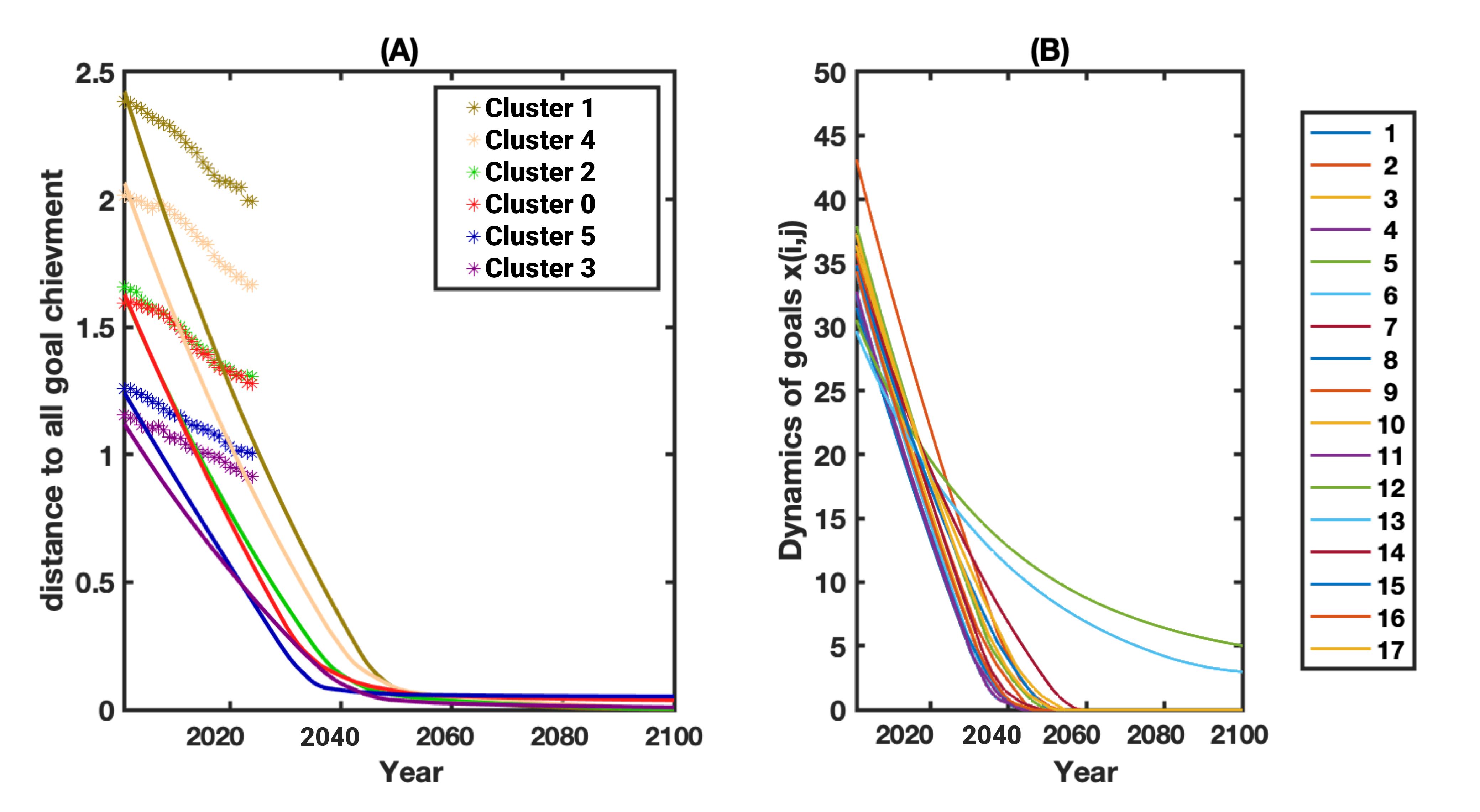}
         \caption{(A) Trajectories of clusters  when the resources are the same for all clusters and the annual spending is 30$\%$ instead of 10$\%$, real data (color stars) and the results of the calculation (color lines). (B) Dynamical behaviour of the 17 SDG goals.}
         \label{fig:r1}
        \label{ra30}
\end{figure}

\AG{It is challenging to advice how a given country should allocate its resources while being devoted to specific goals, due to differences in socio-economic makeups and political demands}. 
However, we could propose a simple scenario in which all countries \AG{allocate} 
30\% \ instead of 10\% \ of their resources annually. The results are shown in Fig. \ref{ra30} \AG{demonstrate the dramatic effect that all the SDG goals are quite satisfactorily approaching success around the year 2040}. 

\newpage
\section{Conclusions}
\AG{In this study we have presented a model to describe the dynamics of Sustainable Development Goals of  countries}. 
This model is suitable not only \AG{to reproduce} 
the dynamics of the real data, but also predict dynamical behavior of \AG{SDGs of countries} 
in \AG{various} hypothetical scenarios. Furthermore, the  model enables us to investigate 
the main 
factors that impede or favour progress, namely the cooperation between countries, correlations between goals, and economic issues. In \AG{our previous study Ref.\cite{SDG2025} we had extrapolated the dynamics of SDGs to future situations}, while here we are \AG{using the model to "predict” situations in different circumstances}. 
Another advantage of \AG{the present modelling methodology is that we can} 
analyze the factors and properties under 
large amount of data that AI allows to process.

However, there is a word of warning concerning the validity of the predictions made under different circumstances, which should in principle be free from random or unknown quantities. For example, in our rough treatment of the possible scenarios, we used a random matrix $e(i,\alpha)$, since we lack data on the precise way each country spends its resources to achieve each of the goals. This is one of the reasons why we decided to take averages over the clusters \AG{of similarly behaving countries}, as they should be less sensitive to details. 

\AG{The present mathematical modelling approach could be extended in the future} 
if one discovers that there are more important factors that should be taken into account to simulate the dynamics of real situations in a given locality and on a specific goal. \AG{This study has shown} 
that it would be 
difficult to make 
progress in achieving the 2030 agenda in a reasonable period of time, unless one proposes a more detailed program that takes into account the specific needs, cultural differences \AG{and socio-economics} of each country.

Finally, we would like to highlight the main conclusions of this work. Firstly, the model, although apparently oversimplified, is able to capture the behaviour of the real system over a long time span. Secondly, the model uncovers the main factors that make the SDG's unattainable in the near future. These factors are the resources countries spend on the task, the communication and cooperation between them, the correlation between goals, and the memory of the actions. Furthermore, this kind of modeling approach could be of help in advising people in charge of making decisions to change the course in order to increase the pace to attain the objective. 

\newpage

\section*{Appendices}
\appendix
\counterwithin{figure}{section}
\setcounter{figure}{0}
\section{Definition of SDG}
\label{sec:SDG}
In here we show in a familiar graphical way the description of the SDG's to facilitate their identification if needed.

\begin{figure}[ht]
\centering
\includegraphics[width=0.8\linewidth]{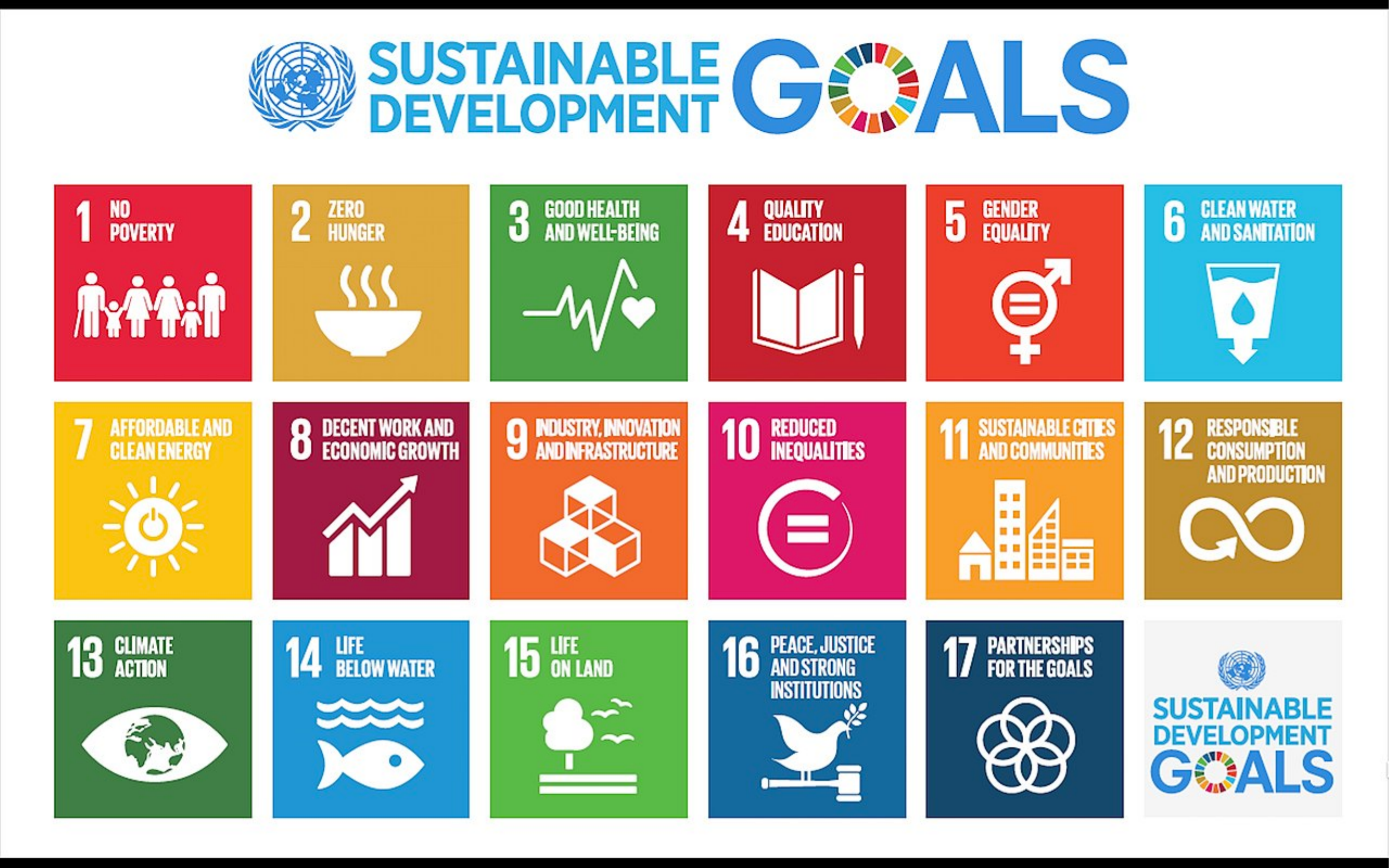}
\caption{Meaning of the 17 SDG according to the UN }
\label{SDG}
\end{figure}

\section{Data sources}
Data were obtained from the Sustainable Development Report 2025~\cite{sdgindex_2024_sustainable}. We used the scores reported for each goal, obtained for each country, evaluated from 2000 to 2024. Goals are subdivided in targets which are evaluated using standard indicators and sub-indicators proposed by the Agenda 2030. SDG evaluation is an integral process that includes compilation and analysis of the indicators at various evaluation levels (policies, programs and projects), as well as various approaches to measure relevance, implementation, progress, and results. All this is done within a framework that promotes equity and respect for national realities.

The clustering used in this work is the one published  previously  (ref. \cite{SDG2025}). Also,  we used real data corresponding to year 2000, for the 107 countries, which are the ones without missing data and data of Gross Domestic Product per capita (GDP/c) in 2022 collected from “Worldometer” [Project RTS. GDP by Country - Worldometer — worldometers.info (ref. 2024\cite{wm}, 
to grossly simulate the amount of resources expended by countries to achieve the SDG’s.

\newpage

\bibliography{references}

\begin{thebibliography}{10}

\bibitem{URL:Goals}
Nations U. {T}{H}{E} 17 {G}{O}{A}{L}{S} | {S}ustainable {D}evelopment ---  sdgs.un.org;.
\newblock \url{https://sdgs.un.org/goals}.

\bibitem{GarcaRodrguez2025}
García-Rodríguez,  Alberto and Núñez,  Matias and Pérez,  Miguel Robles and Govezensky,  Tzipe and Barrio,  Rafael A and Gershenson,  Carlos and Kaski,  Kimmo K and Tag\"{u}eña,  Julia.
\newblock {Sustainable visions: unsupervised machine learning insights on global development goals}.
\newblock   PLOS ONE.
\newblock doi:{10.1371/journal.pone.0317412}
\newblock Available from: \url{http://dx.doi.org/10.1371/journal.pone.0317412}
\end{thebibliography}

\end{document}